\documentclass[3p,times]{elsarticle}

\usepackage{ecrc}

\volume{00}

\firstpage{1}

\journalname{Nuclear Physics A}

\runauth{C. Shen et al.}

\jid{npa}

\jnltitlelogo{Nuclear Physics A}

\usepackage{amssymb}
\usepackage{mathrsfs}
\usepackage{amsfonts}
\usepackage{amsthm}
\usepackage{latexsym}
\usepackage{slashed}

\biboptions{square,comma,numbers,sort&compress}

\usepackage[figuresright]{rotating}

\begin{document}

\begin{frontmatter}



\dochead{}

\title{Thermal photon anisotropic flow serves as a quark-gluon plasma viscometer}

\author[label1]{Chun Shen}
\author[label1]{Ulrich Heinz}
\author[label2]{Jean-Fran\c{c}ois Paquet}
\author[label2]{Charles Gale}

\address[label1]{Department of Physics, The Ohio State University, Columbus, Ohio 43210-1117, USA}
\address[label2]{Department of Physics, McGill University, 3600 University Street, Montreal, Quebec, H3A 2T8, Canada}

\begin{abstract}
Photons are a penetrating probe of the hot and dense medium created in heavy-ion collisions. We present state-of-the-art calculations of viscous photon emission from nuclear collisions at RHIC and LHC. Thermal photons'\ anisotropic flow coefficients $v_n$ are computed, both with and without accounting for viscous corrections to the standard thermal emission rates. These corrections are found to have a larger effect on the $v_n$ coefficients than the viscous suppression of hydrodynamic flow anisotropies. For thermal photons, the ratio $v_2\{\mathrm{SP}\}/v_3\{\mathrm{SP}\}$ shows stronger sensitivity to the quark-gluon plasma (QGP) shear viscosity than for hadrons, and it can thus serve as a sensitive QGP viscometer.
\end{abstract}


\end{frontmatter}


\section{Introduction}
\label{intro}

Heavy-ion collisions at the Relativistic Heavy-Ion Collider (RHIC) and the Large Hadron Collider (LHC) allow to study the physics of strongly interacting quark-gluon plasma (QGP) under conditions of extreme heat and density. Electromagnetic radiation is a sensitive and direct probe of the space-time evolution in heavy-ion collisions. Thermal photons provide information about the early dynamics of heavy-ion collisions that is complementary to that obtained from the wealth of hadronic observables. Recent measurements show a surprisingly large second order azimuthal anisotropy of direct photons, comparable with the elliptic flow of hadrons, both in Au+Au collisions at RHIC \cite{Adare:2011zr,Tserruya:2012jb}
and Pb+Pb collisions at the LHC \cite{Lohner:2012ct}. These results challenge our current theoretical understanding of thermal photon production \cite{Chatterjee:2005de}. In this work, we present state-of-the-art calculations of thermal photon anisotropic flows from event-by-event viscous hydrodynamic simulations \cite{Shen:2013cca}. We find that thermal photon anisotropic flows are more sensitive to the shear viscosity of the medium than their charged hadron analogues. The ratio of elliptic and triangular flow of direct photons is shown to serve as a sensitive QGP viscometer. 

\section{Methodology}
We employ the boost-invariant viscous hydrodynamic code {\tt VISH2+1} which has been very successful in describing soft hadron observables at RHIC and LHC energies \cite{Shen:2011eg,Qiu:2011hf}. Initial conditions are generated from the Monte-Carlo Glauber (MCGlb) and Monte-Carlo KLN (MCKLN) models and then evolved on an event-by-event basis through viscous hydrodynamics, using a lattice-based equation of state with chemical freeze-out at $T_\mathrm{chem}=165$\,MeV \cite{Huovinen:2009yb}. The hydrodynamic parameters were determined in \cite{Shen:2011eg,Qiu:2011hf} by fitting hadronic observables at both RHIC and LHC energies. We start the hydrodynamic simulations at $\tau_0 = 0.6$\,fm/$c$ and decouple at $T = 120$\,MeV. For MCGlb initial conditions, we use the specific shear viscosity $\eta/s = 0.08$ while MCKLN initial conditions require  $\eta/s = 0.20$ \cite{Shen:2011eg}. The thermal photon momentum distribution can be calculated by folding the thermal emission rate with the evolving hydrodynamic medium: 
\begin{equation}
E \frac{dN^\gamma}{d^3 p} = \int \tau d\tau dx dy d\eta \left(\Gamma_0(q, T) + \frac{q_\mu q_\nu \pi^{\mu\nu}(x)}{2(e(x){+}P(x))}a_{\alpha \beta} \Gamma^{\alpha \beta}(q, T) \right) \bigg\vert_{q = p \cdot u(x); T(x)}.
\label{eq1}
\end{equation}
The photon emission rate includes a dominant equilibrium contribution, $\Gamma_0(q, T)$, and a first order viscous correction proportional to the shear stress tensor, $\frac{\pi^{\mu\nu}q_\mu q_\nu}{2(e+p)}a_{\alpha \beta} \Gamma^{\alpha \beta}(q, T)$, where $a_{\mu \nu} = \frac{3 q_\mu q_\nu}{2(u\cdot q)^4} + \frac{u_\mu u_\nu}{(u \cdot q)^2} + \frac{g_{\mu\nu}}{2(u \cdot q)^2} - \frac{3(q_\mu u_\nu + q_\nu u_\mu)}{2(u \cdot q)^3}$ \cite{Shen:2013cca}. The rates $\Gamma_0(q, T)$ and $\Gamma^{\alpha \beta}(q, T)$ are calculated using local distribution functions with viscous corrections linear in the shear stress, $f(x,p) = f_0(x,p) \left[1 + (1{\pm}f_0(x,p)) \frac{p_\mu p_\nu\,\pi^{\mu\nu}(x)}{2T^2(x)(e(x){+}P(x))}\right]$. The final thermal photon anisotropic flow coefficients are calculated using the photon multiplicity weighted scalar product method:
\begin{equation}
 v^\gamma_{n} \{ \mathrm{SP} \} ( p_{T} )  =  \frac{\left\langle \frac{d
  N^{\gamma}}{d y p_{T} d p_{T}} ( p_{T} ) v_{n}^{\gamma} ( p_{T} )
  v_{n}^{\mathrm{ch}} \cos ( n ( \Psi_{n}^{\gamma} ( p_{T} ) -
  \Psi_{n}^{\mathrm{ch}} ) ) \right\rangle}{\left \langle \frac{d N^{\gamma}}{d y
  p_{T} dp_{T}} ( p_{T} ) \right\rangle v_{n}^{\mathrm{ch}} \{ 2 \}}.
\label{eq2}
\end{equation}
Here $\langle ... \rangle$ represents an average over events, and $v_n^\mathrm{ch}$ and $\Psi_n^\mathrm{ch}$ are the $p_T$-integrated charged hadron anisotropic flow coefficients and their associated flow plane angles in each event. $v_{n}^{\mathrm{ch}} \{ 2 \}$ is the charged hadron rms flow of order $n$, extracted from two-particle cumulants.

\section{Results}
\label{results}
%
\begin{figure}
\begin{tabular}{cc}
  \includegraphics[width=0.45\linewidth]{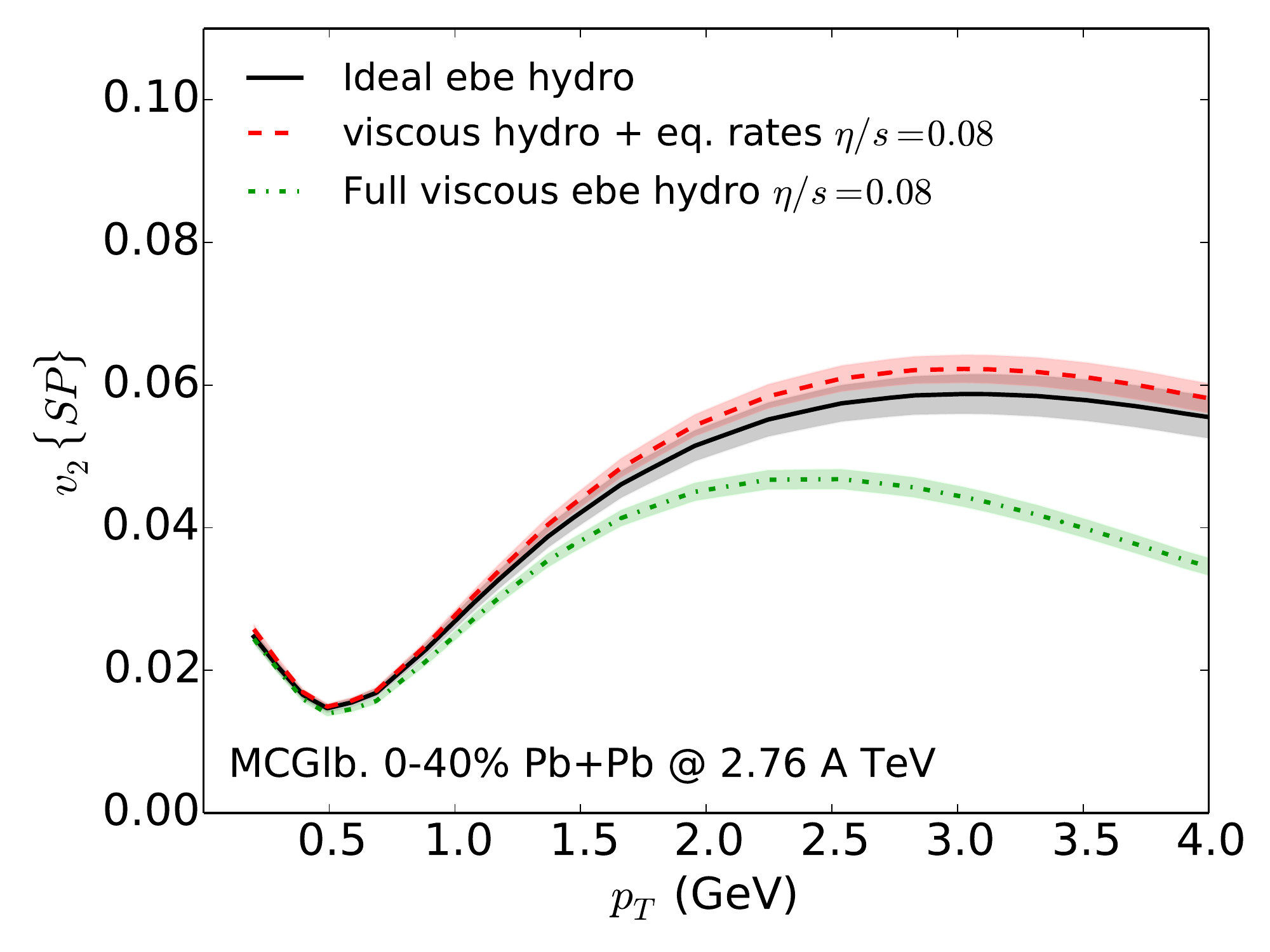} & 
  \includegraphics[width=0.45\linewidth]{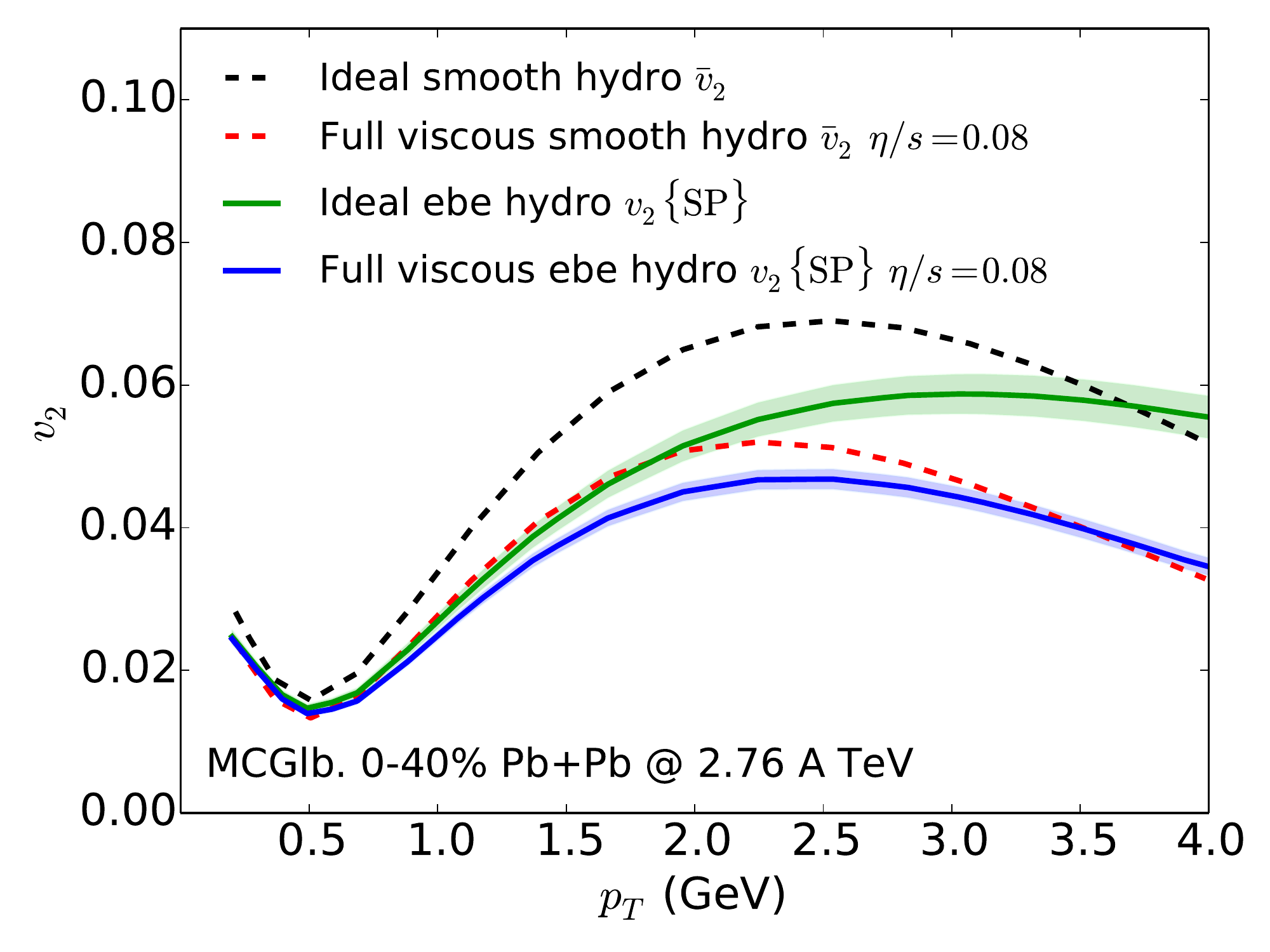}
\end{tabular}
\caption{Thermal photon elliptic flow as a function of $p_T$ at 0-40\% centrality Pb+Pb collisions at $\sqrt{s} = 2.76$ $A$ TeV. Left panel: Comparisons between thermal photons emitted from ideal and viscous hydrodynamic medium. Right panel: Comparisons between thermal photons elliptic flow from smooth event averaged and event-by-event hydrodynamic medium.}
\label{fig1}
\end{figure}
%
In the left panel of Fig.~\ref{fig1}, we study the shear viscous correction to the thermal photon elliptic flow. Compared to results from an ideal hydrodynamic medium, the viscous hydrodynamic evolution alone results in slightly larger $v_2$ at high $p_T$. This is because the initial temperature of the viscous hydrodynamic medium is lower than in the ideal case, in order to compensate for viscous entropy production during the evolution. This reduces the proportion of high momentum photons coming from the early hot region of the fireball which carries small elliptic flow. However, this increase in elliptic flow is overwhelmed by a much larger suppression arising from the viscous correction to the thermal photon emission rate. In contrast to the elliptic flow of hadrons, the viscous correction to the photon emission rates dominates the suppression for photons. In the right panel of Fig. \ref{fig1}, we illustrate the difference between thermal photon $v_2$ from smooth event-averaged and event-by-event hydrodynamic calculations. For both ideal and viscous cases, the photon $v_2\{\mathrm{SP}\}$ from event-by-event simulations are smaller than the mean $v_2$ from the smooth event-averaged hydrodynamic medium. Although the initial state fluctuation increases the final flow anisotropy, the photon multiplicity factor in Eq.~(\ref{eq2}) biases the measurement towards central collisions with smaller elliptic flow, especially for bumpy profiles with temperature hot spots. 

\begin{figure}
\begin{tabular}{cc}
  \includegraphics[width=0.45\linewidth]{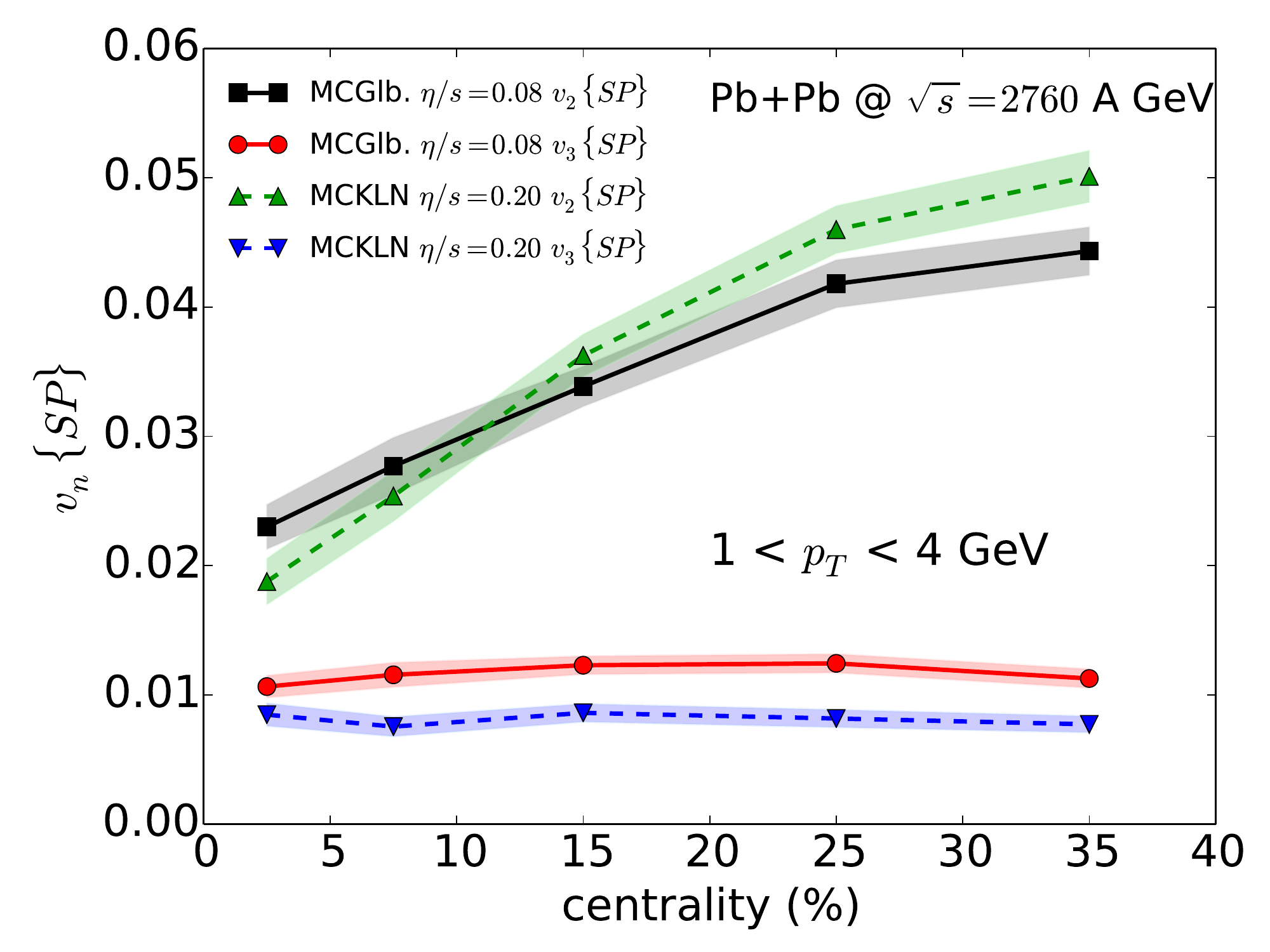} & 
  \includegraphics[width=0.45\linewidth]{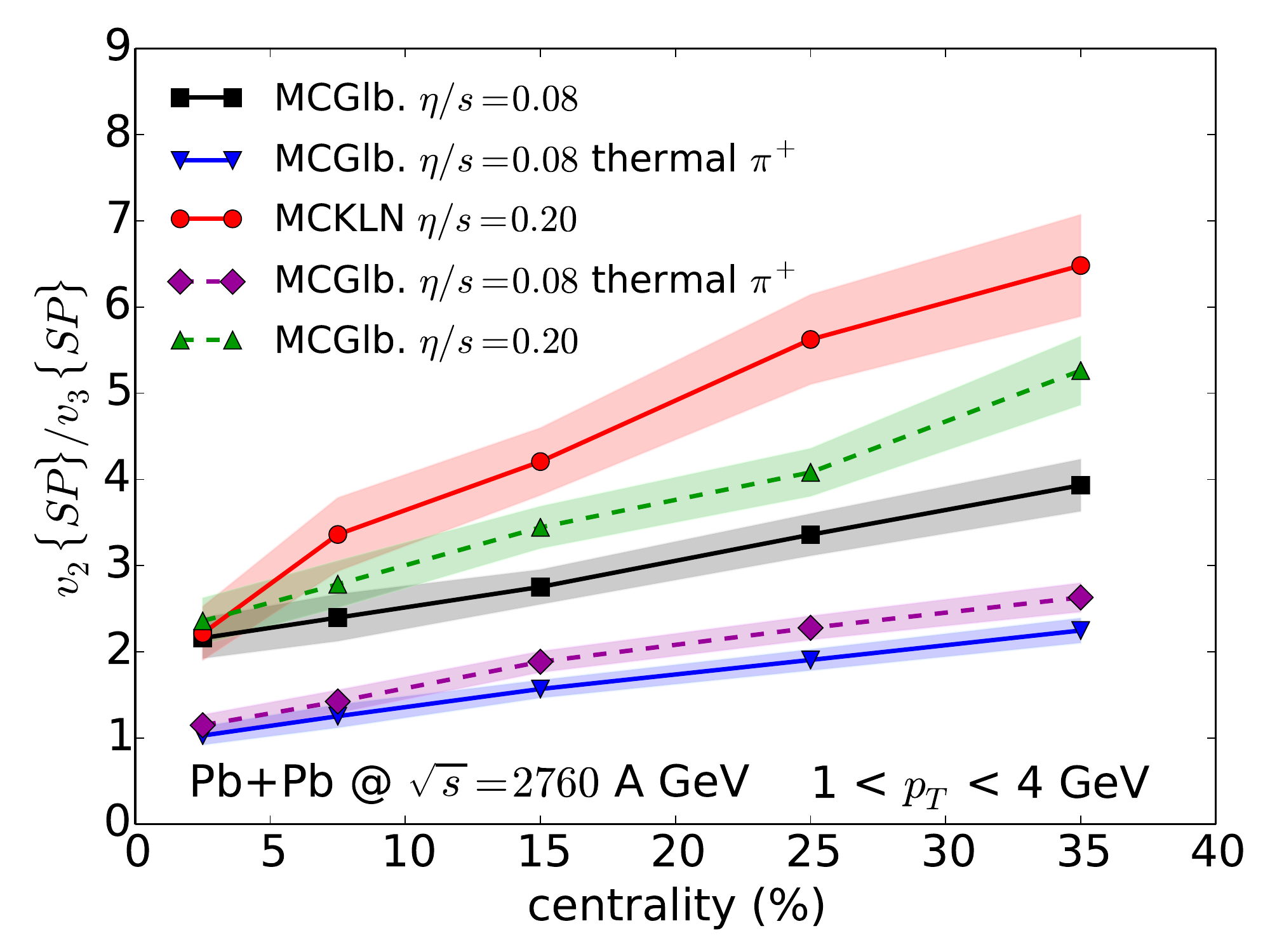}
\end{tabular}
\caption{Left panel: Centrality dependence of the $p_T$-integrated $v_n\{\mathrm{SP}\}$ of thermal photons from MCGlb. and MCKLN initial conditions. $p_T$ is integrated from 1 to 4 GeV. Right panel: The corresponding ratio $v_2\{\mathrm{SP}\}/v_3\{\mathrm{SP}\}$ as a function of centrality. }
\label{fig2}
\end{figure}
%
The centrality dependence of the $p_T$-integrated thermal photon $v_n\{\mathrm{SP}\}$ is shown in the left panel of Fig.~\ref{fig2} for $n = 2, 3$. The elliptic flow of thermal photons increases with centrality due to the increasing geometric eccentricity of the nuclear overlap area. The triangular flow has little centrality dependence because it is purely driven by event-by-event fluctuations. The ratio between thermal photon elliptic and triangular flow is shown in the right panel of Fig.~\ref{fig2}. Since prompt photons carry a small anisotropy their contribution is expected to be negligible in this ratio:
\begin{equation}
  \frac{v^\gamma_{2} \{ \mathrm{SP} \} ( p_{T} )}{v^\gamma_{3} \{ \mathrm{SP} \} ( p_{T} )} \simeq
  \frac{\left\langle \frac{d N^{\gamma , \mathrm{thermal}}}{d y p_{T}
  \mathrm{dp}_{T}} ( p_{T} ) v_{2}^{\gamma , \mathrm{thermal}} ( p_{T} )
  v_{2}^{\mathrm{ch}} \cos ( 2 ( \Psi_{2}^{\gamma , \mathrm{thermal}} ( p_{T} ) -
  \Psi_{2}^{\mathrm{ch}} ) ) \right\rangle}{\left\langle \frac{d N^{\gamma ,
  \mathrm{thermal}}}{d y p_{T} \mathrm{dp}_{T}} ( p_{T} ) v_{3}^{\gamma ,
  \mathrm{thermal}} ( p_{T} ) v_{3}^{\mathrm{ch}} \cos ( 3 ( \Psi_{3}^{\gamma ,
  \mathrm{thermal}} ( p_{T} ) - \Psi_{3}^{\mathrm{ch}} ) ) \right\rangle}  
  \frac{v_{3}^{\mathrm{ch}} \{ 2 \}}{v_{2}^{\mathrm{ch}} \{ 2 \}}. 
\label{eq3}
\end{equation}
The ratio $v^\gamma_{2} \{ \mathrm{SP} \} /v^\gamma_{3} \{ \mathrm{SP} \}$ increases with the specific shear viscosity $\eta/s$ of the medium, reflecting the fact that higher order harmonic flows are suppressed more strongly by viscous effects. Comparing MCGlb results for different $\eta/s$, we find that for photons this ratio has a larger sensitivity to shear viscosity than the same ratio for thermal pions \cite{Shen:2013cca}. For fixed $\eta/s = 0.20$, the centrality dependence of this ratio is stronger for MCKLN than for MCGlb initial conditions. This is due to the stronger centrality dependence of  $v^\gamma_{2} \{ \mathrm{SP} \} $ in the MCKLN model, shown in the left panel of Fig.~\ref{fig2}. 

\begin{figure}
\begin{tabular}{cc}
  \includegraphics[width=0.45\linewidth]{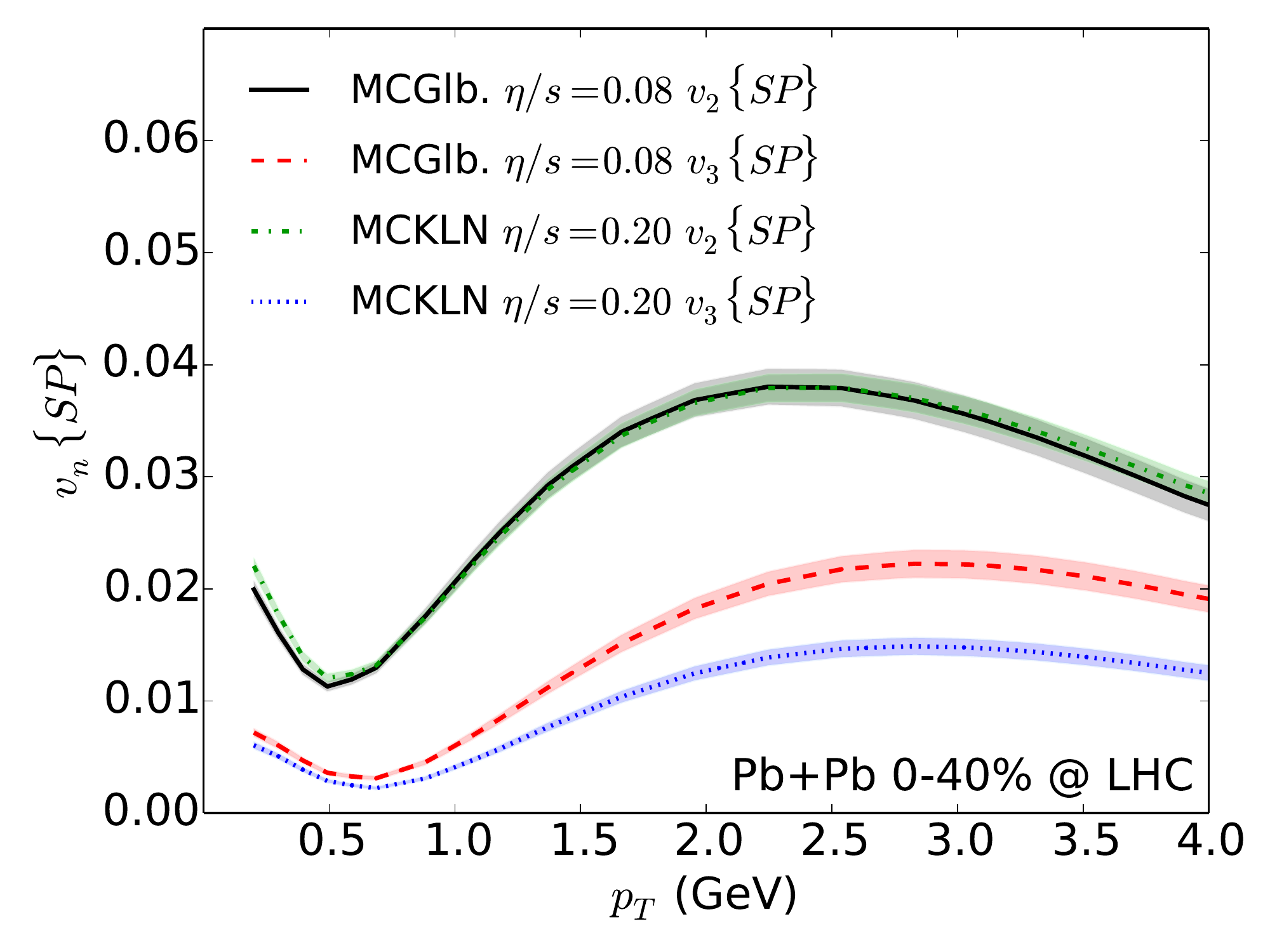} & 
  \includegraphics[width=0.45\linewidth]{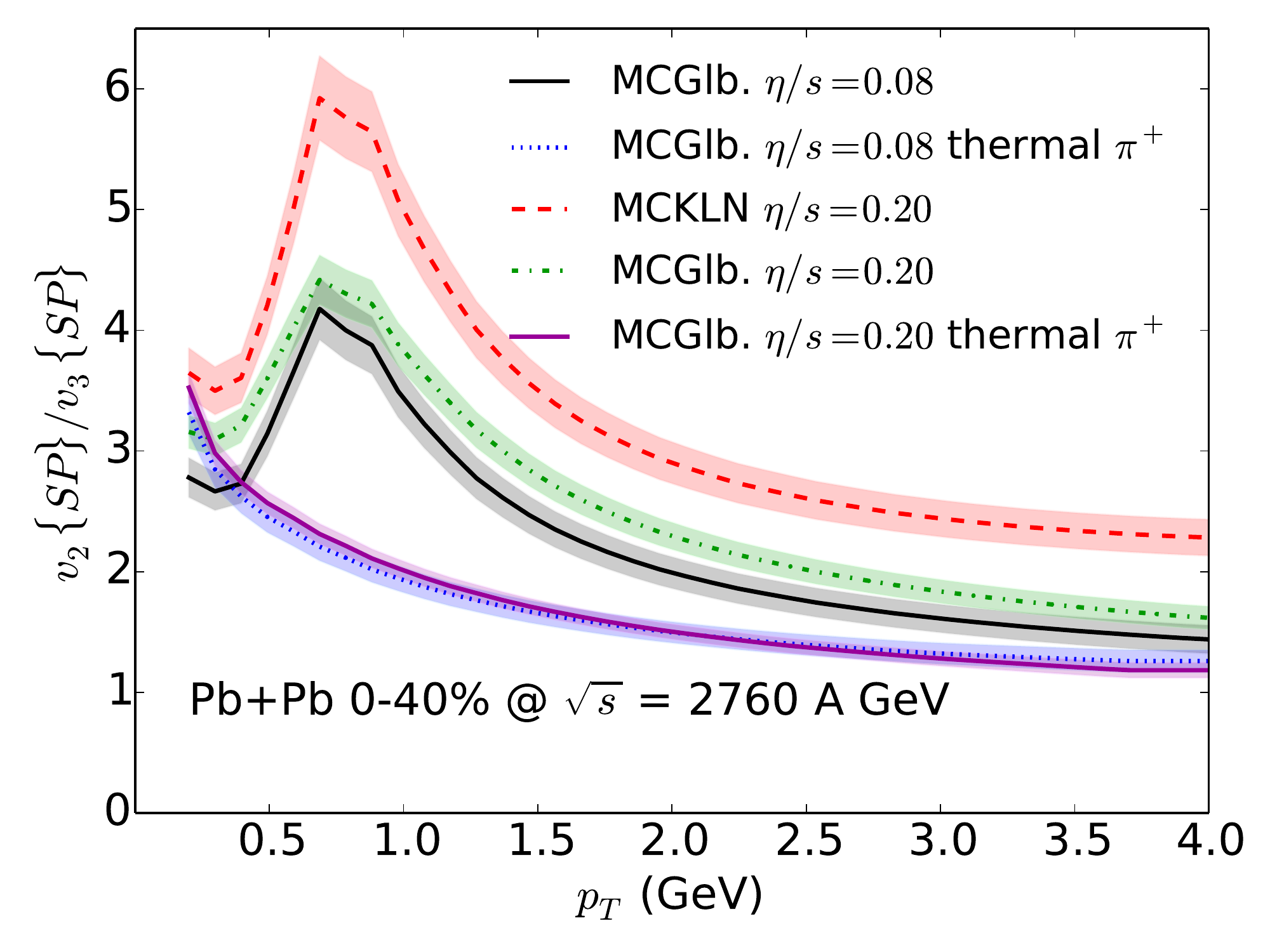}
\end{tabular}
\caption{Left panel: $p_T$-differential $v_{2.3}\{\mathrm{SP}\}$ of thermal photons at 0-40\% centrality in Pb + Pb collisions at $\sqrt{s} = 2.76$ $A$ TeV.  Right panel: The corresponding ratio $v_2\{\mathrm{SP}\}/v_3\{\mathrm{SP}\}$ as a function of $p_T$ compared with the same ratio for thermal $\pi^+$. }
\label{fig3}
\end{figure}

Proceeding to Fig.~\ref{fig3}, we show in the left panel the $p_T$-differential elliptic and triangular flows of thermal photons. The two initial condition models (with their correspondingly adjusted choices $\eta/s$) produce very similar thermal photon elliptic flows, $v^\gamma_{2} \{ \mathrm{SP} \}$, whereas
the triangular flow from MCKLN initial conditions is only about half as large as for the MCGlb model, due to the larger $\eta/s$ used in evolving the MCKLN profiles. Compared to hadrons, the thermal photon flow coefficients $v_n\{\mathrm{SP}\}(p_T)$ have a richer structure. They decrease 
%
at $p_T \gtrsim 2.5$ GeV due to the increasing weight of photons coming from early times when  hydrodynamic flow is still weak \cite{Chatterjee:2005de}. The dip at $p_T$ around the mass of  the $\rho$ meson is due to the transition from $\pi{\,+\,}\pi{\,\rightarrow\,}\rho{\,+\,}\gamma$ to $\pi{\,+\,}\rho{\,\rightarrow\,}\pi{\,+\,}\gamma$ as the dominant photon production channel as $p_T$ increases \cite{Chatterjee:2005de}. Since for the hadronic mesons, who transfer their flow to the photons, $v_n(p_T){\,\sim\,}p^n_T$ at small $p_T$, this dip shifts slightly towards larger $p_T$ as $n$ increases. This small shift generates an interesting rise-and-fall structure in the low-$p_T$ region for the ratio $v^\gamma_{2} \{ \mathrm{SP} \}(p_T) /v^\gamma_{3} \{ \mathrm{SP} \}(p_T)$, shown in the right panel of Fig.~\ref{fig3}. Compared to charged hadrons, the thermal photon $v^\gamma_{2} \{ \mathrm{SP} \} /v^\gamma_{3} \{ \mathrm{SP} \}$ ratio shows higher sensitivity to the shear viscosity of the radiating medium. We found that this sensitivity is even stronger for the ratios $v^\gamma_{2} \{ \mathrm{SP} \} /v^\gamma_{n} \{ \mathrm{SP} \}$ with $n \ge 4$ (not shown). 

Equation~(\ref{eq3}) shows that the $v^\gamma_{2} \{ \mathrm{SP} \} /v^\gamma_{3} \{ \mathrm{SP} \}$ ratio in Figs.\,\ref{fig2} and \ref{fig3} is insensitive to the yields from direct photon sources that carry zero anisotropic flow. A measurement of this ratio thus allows to focus attention on those photons that reflect and probe the space-time history of the evolving near-thermal medium. Pre-equilibrium photons generated by the strong magnetic field of the passing nuclei through the QCD anomaly \cite{Basar:2012bp} are predicted to carry sizable $v_2$ but are expected to have small triangular flow $v_3$; they would generate much larger $v_2/v_3$ than thermal photons. Measuring this ratio will allow to disentangle these two mechanisms for the already observed (large) direct photon elliptic flow. 

\section{Conclusions}
\label{conclusion}
State-of-the-art calculations of thermal photon anisotropic flow, $v_n\{\mathrm{SP}\}$ ($n = 2,3$), use event-by-event viscous hydrodynamic simulations to account for event-by-event quantum fluctuations in the initial state. Shear viscosity suppresses photon $v_n\{\mathrm{SP}\}$, with viscous corrections to the photon production rates dominating this suppression. For both the $p_T$-integrated and $p_T$-differential anisotropic flows, the ratio $v^\gamma_{2} \{ \mathrm{SP} \} /v^\gamma_{3} \{ \mathrm{SP} \}$ shows stronger sensitivity to the specific shear viscosity of the QGP for thermal photons than for charged hadrons. This ratio increases with $\eta/s$ because the viscous suppression of $v_n$ increases with the harmonic order $n$. Since the ratio $v^\gamma_{2} \{ \mathrm{SP} \} /v^\gamma_{3} \{ \mathrm{SP} \}$ is insensitive to photon sources that carry zero anisotropic flow, such as prompt photons, the experimental measurements of this ratio for direct photons will shed new and more direct light on the specific shear viscosity of the thermal medium formed after the end of prompt photon emission, but well before most of finally emitted hadrons are set free.

\bigskip
\noindent
{\bf Acknowledgments: }
This work was supported by the U.S. Department of Energy under Grants No. \rm{DE-SC0004286} and (within the framework of the JET Collaboration) \rm{DE-SC0004104} and by the Natural Sciences and Engineering Research Council of Canada.
%




\bibliographystyle{elsarticle-num}
\bibliography{photon_vn_ref.bib}







\end{document}